# Automatic Identification of the End-Diastolic and End-Systolic Cardiac Frames from Invasive Coronary Angiography Videos

Yinghui Meng[1], Minghao Dong[1], Xumin Dai[4], Haipeng Tang[2], Chen Zhao[3], Jingfeng Jiang[3], Shun Xu[5], Ying Zhou[5], Fubao Zhu[1], Zhihui Xu[5*], Weihua Zhou[3,6*]

[1]School of Computer and Communication Engineering, Zhengzhou University of Light Industry, Zhengzhou, Henan, China

[2]School of Computing Sciences and Computer Engineering, University of Southern Mississippi, Hattiesburg, MS, USA

[3]Department of Applied Computing, Michigan Technological University, Houghton, MI, USA

[4]Department of Cardiology, Theresa & Eugene M. Lang Center for Ressearch & Education, New York Presbyterian Queens Hospital, New York, USA

[5]Department of Cardiology, The First Affiliated Hospital of Nanjing Medical University, Nanjing, China

[6]Center for Biocomputing and Digital Health, Institute of Computing and Cybersystems, and Health Research Institute, Michigan Technological University, Houghton, MI, USA

*Corresponding Authors:

Weihua Zhou, PhD E-Mail: whzhou@mtu.edu
Department of Applied Computing, Michigan Technological University
1400 Townsend Dr, Houghton, MI, 49931, USA
Tel: 906-487-2666

Or

Zhihui Xu, MD E-mail: wx_xzh@njmu.edu.cn
Department of Cardiology, The First Affiliated Hospital of Nanjing Medical University,
300 Guangzhou Rd, Gulou, Nanjing, 210000, China;
Tel: (+86)02568303120

**Abstract:** Automatic identification of proper image frames at the end-diastolic (ED) and end-systolic (ES) frames during the review of invasive coronary angiograms (ICA) is important to assess blood flow during a cardiac cycle, reconstruct the 3D arterial anatomy from bi-planar views, and generate the complementary fusion map with myocardial images. The current identification method primarily relies on visual interpretation, making it not only time-consuming but also less reproducible. In this paper, we propose a new method to automatically identify angiographic image frames associated with the ED and ES cardiac phases by using the trajectories of key vessel points (*i.e.* landmarks). More specifically, a detection algorithm is first used to detect the key points of coronary arteries, and then an optical flow method is employed to track the trajectories of the selected key points. The ED and ES frames are identified based on all these trajectories. Our method was tested with 62 ICA videos from two separate medical centers (22 and 9 patients in sites 1 and 2, respectively). Comparing consensus interpretations by two human expert readers, excellent agreement was achieved by the proposed algorithm: the agreement rates within a one-frame range were 92.99% and 92.73% for the automatic identification of the ED and ES image frames, respectively. In conclusion, the proposed automated method showed great potential for being an integral part of automated ICA image analysis.

## 1. Introduction

Coronary artery disease (CAD) is a lethal disease posing threats to human health [1]. Myocardial revascularization (MR) is often a necessary interventional treatment for patients with symptomatic CAD [2]. In addition to clinical symptomology, the degree or percentage of stenosis in major epicardial coronary arteries are often used to guide MR decision-making. Invasive coronary angiography (ICA) is the most commonly used imaging modality for the assessment of the anatomical structure and severity of stenosis within coronary arteries [3]. The principle of ICA is to use the X-ray cine angiography to visualize the coronary arteries by selectively injecting radial opaque contrast medium into coronary arteries [4]. By visually inspecting the 2-D ICA images, physicians then determine whether there is a presence of a stenosis, and the location and severity of the lesion given the presence of the stenosis, and make clinical decisions on the choice of treatment approaches.

ICA is often acquired as videos in which time-resolved two-dimensional arteries are presented; the motion of arteries in those videos can assist in resolving the overlaps and assessing blood flow during a cardiac cycle. At the same time, three-dimensional reconstruction of the arterial anatomy for bi-planar ICA views further facilitates the visualization [5],[6]. However, clinical trials showed that, for the treatment of patients with stable CAD, aggressive MR solely based on the percentage of stenosis yielded no better mortality outcome than the use of optimal medical therapy [7],[8],[9]. Adding myocardial function information, obtained through advanced myocardial imaging, such as SPECT myocardial perfusion imaging and cardiac magnetic resonance imaging (CMR), to ICA anatomical information, in assisting the selection of patients with stable CAD for MR treatment, was found to improve outcomes [10].

In both 3D reconstructions of arteries and multi-modality image fusion, automatic identification of matched end-diastole (ED) and end-systole (ES) image frames from different ICA views is critical. For instance, both flow fraction reserve (FFR) calculated from the 3D arterial model [11] and 3D quantitative coronary angiography (QCA) [12] require ED image frames in ICA videos. The identification of ES image frames is essential for the diagnosis of the myocardial bridge [13]. The ideal technique for extracting ED and ES from ICA images is to display electrocardiograms (ECG) on ICA in real-time, whereas the module has not been standardized and is often missing in the clinical workflow.

In this paper, we propose a novel approach for the automatic identification of ED and ES frames through a sequence of ICA videos. Our approach first uses a detection algorithm to detect the key points of coronary arteries and then an optical flow method to track the trajectories of the selected key points. The ED and ES frames are identified based on all these trajectories. To evaluate our proposed method, we retrospectively investigated 31 patients from two medical centers. The identified ED and ES frames were compared with the ground truth, which are consensus readings by physicians.

## 2. Materials and Method

**Patients Data**

This study includes de-identified ICAs of 31 patients (24 males) from two clinical centers. Twenty-two patients are enrolled from Jiangsu Provincial People's Hospital (Site 1) and the rest from Queens Hospital in New York (Site 2). The exemption status of the study protocol was granted by respective local institutional review board. Table 1 lists three interventional angiography systems used in two hospitals for fluoroscopy angiography and the acquisition speed is 15 frames/sec. Three image sizes (512×512, 864×864, 1000×1000) are used. Each patient has two DICOM videos from two image acquisition views.

**Table 1.** Image size and the number of patients collected by different interventional angiography systems in two participating hospitals.

| | Manufacturer /Manufacturer's Model Name | Image size | Number of patients |
|---|---|---|---|
| Site 1 | Siemens/AXIOM-Artis | 512×512 | 9 |
| | GE MEDICAL SYSTEMS/ DL | 864×864 | 10 |
| | | 1000×1000 | 1 |
| | Philips Medical Systems/ AlluraXper | 512×512 | 2 |
| Site 2 | Philips Medical Systems/ AlluraXper | 512×512 | 9 |

**Procedures**

Our approach has three major steps: first, key points of coronary arteries are selected from the first frame of the ICA video with minimal human input. Left anterior descending artery (LAD) and left coronary artery (LCX) are shown in this frame; second, the Lucas-Kanade (LK) pyramid optical flow method is used to track the trajectories of the selected key points; finally, the numbers of ED and ES frames are obtained from the trajectory diagrams.

**Selection of key points**

Since end-diastolic and end-systolic frames are identified based on the trajectories of trackable vessels, the ideal key points are the intersections of blood vessels. In order to eliminate the distraction of non-vessel objects, such as electrodes (Figure 1B), a rectangular region of interest (ROI) is drawn by operators to select the arterial territory of interest, as illustrated in Figure 1C.

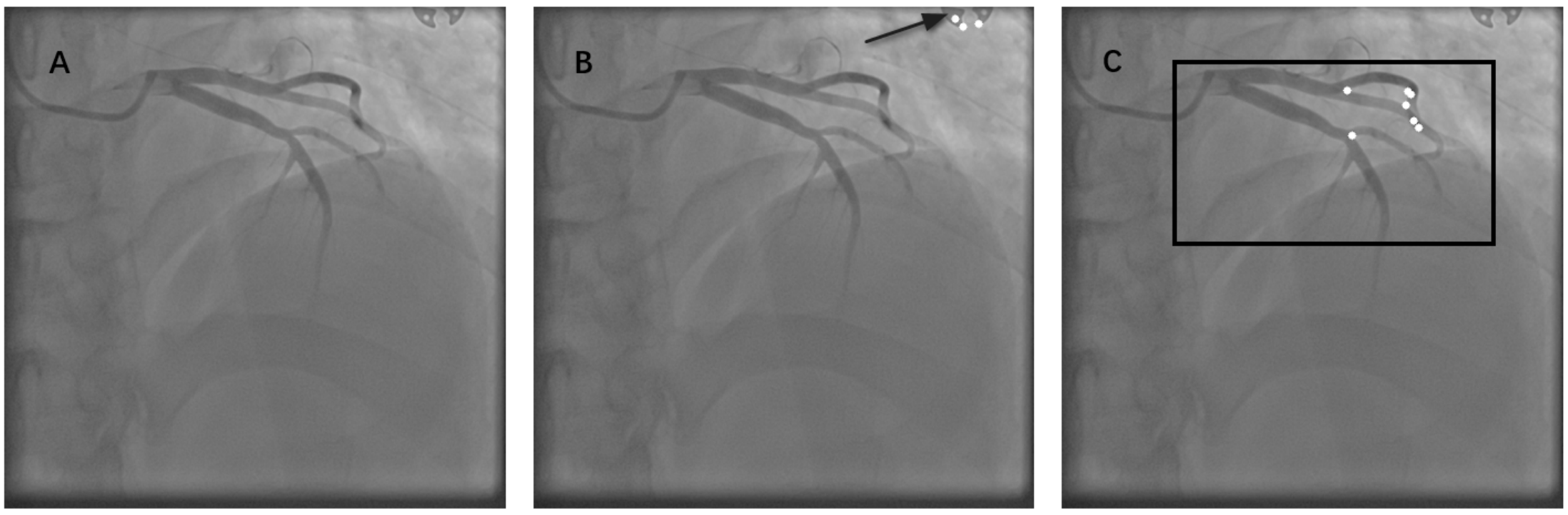

**Figure 1**. Detection of key points from the first frame. (A) The first frame, selected by operators for a video of invasive coronary angiography. (B) Non-vessel objects (electrodes by arrow in this figure), which could interfere with the selection of key points. (C) A rectangular region of interest, drawn by an operator to guide the selection of key points.

Once the ROI is selected, a Shi-Tomasi corner detection algorithm [15] is adopted to automatically identify the key points. This algorithm calculates the associated position where the local image intensity changes dramatically by using a sliding window with fixed dimensions. The grayscale change of pixels in the corresponding window before and after a shift (u, v) can be expressed as follows:

$$E(u,v)=\sum_{x,y} w(x,y)[I(x+u,y+v)-I(x,y)]^2 \qquad \text{Eq. 1}$$

where $(x,y)$ is the corresponding pixel coordinate of the center position in the window, $I(x,y)$ is the pixel value of the current position, $I(x+u,y+v)$ is the pixel value of the new position after the window is moved $[u,v]$, and $w(x,y)$ is the weight of each pixel in the window.

Using Taylor expansion and a Sobel operator, Eq. 1 can be reorganized as follows:

$$E(u,v)\cong[u,v]M\begin{bmatrix}u\\v\end{bmatrix} \qquad \text{Eq. 2}$$

where M is a 2*2 matrix, $\lambda_1$ and $\lambda_2$ are the rates of gray level changes in the x and y directions, respectively.

$$M=\sum_{x,y} w(x,y)\begin{bmatrix}I_x^2 & I_xI_y\\ I_xI_y & I_y^2\end{bmatrix} \xrightarrow{\text{Diagonalization}} M=\begin{bmatrix}\lambda_1 & 0\\ 0 & \lambda_2\end{bmatrix} \qquad \text{Eq. 3}$$

The Shi-Tomasi algorithm proposes that a strong key point can be obtained if the smaller eigenvalues in the two eigenvalues are greater than the minimum threshold. $R=\min(\lambda_1,\lambda_2)$ is the corresponding response function of key points.

**Motion tracking of the selected key points**

The well-known LK pyramid optical flow algorithm [17] is used to track the motion of key points and estimate their trajectories. Thus, the cardiac phase of the frames can be predicted accordingly. This algorithm builds a Gaussian pyramid from the raw frame image. The bottom is the high-resolution low-level image, and the top is the low-resolution high-level image. Each layer in the pyramid is downsampled from the previous layer to obtain a series of reduced-size images.

Let $I^0 = I$ be the raw image. Then the pyramid is built recursively, with $I^0$ being downsampled to get $I^1$, $I^1$ being downsampled to get $I^2$, and so on $I^{L_m}$ being the top-level image. $L$ is the layer of the pyramid and $I^L$ is the image pyramid.

Eq. 4 defines the coordinate of the key point $p$ obtained from the $L_{th}$ level image.

$$p^L = \frac{p}{2^L} \qquad \text{Eq. 4}$$

The algorithm suggests that the pyramid level should not exceed four levels in most cases. Therefore, a three-level image pyramid was created from adjacent frames of the ICA video.

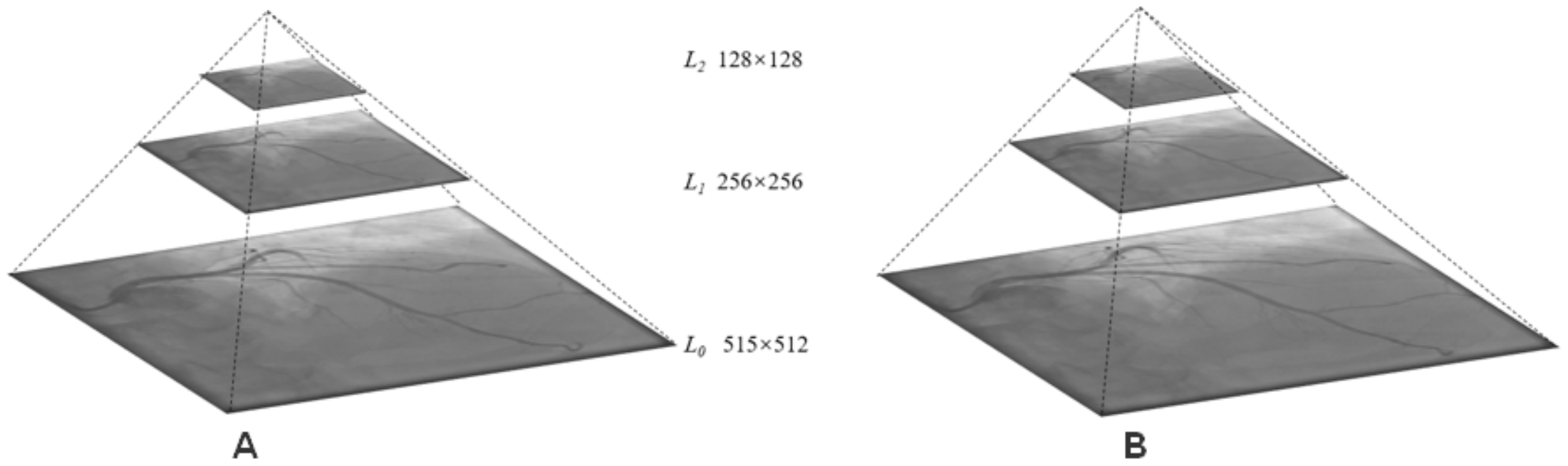

Figure 2. Three levels image pyramid. A and B are two adjacent frames.

For the top-level $L_m$, initial optical flow (vector) field was initialized as zeros:

$$g^{L_m} = \begin{bmatrix} g_x^{L_m} & g_y^{L_m} \end{bmatrix}^T = \begin{bmatrix} 0 & 0 \end{bmatrix}^T \qquad \text{Eq. 5}$$

The optical flow vector $d^{L_m} = \begin{bmatrix} d_x^{L_m} & d_y^{L_m} \end{bmatrix}^T$ at the top pyramid level $L_m$ is computed by minimizing the error function $\varepsilon^L$, as shown in Eq. 6,

$$\varepsilon^L(d^L) = \varepsilon(d_x^L, d_y^L) = \sum_{x=p_x^L-\omega_x}^{p_x^L+\omega_x} \sum_{y=p_y^L-\omega_y}^{p_y^L+\omega_y} (A^L(x,y) - B^L(x+g_x^L+d_x^L, y+g_y^L+d_y^L))^2 \qquad \text{Eq. 6}$$

where $\begin{bmatrix} p_x^L & p_y^L \end{bmatrix}^T$ is the coordinate of the point $p^L$, $w_x$ and $w_y$ are two integers, $A$ and $B$ are the grayscale values at the coordinates. As a result, the estimation of optical flow vector is done over a fixed size kernel $(2w_x + 1, 2w_y + 1)$ centered at $p^L$.

Minimization of Eq. 6 leads to $\frac{\partial \varepsilon^L}{\partial d} = 0$. The similarity between the corresponding points of the two frames of images is the highest.

Then, the optical flow vector corresponding to the top-level will be upsampled by a factor of 2 and fed to the the level $L_m - 1$ as the initial optical flow estimation $g^{L_m - 1}$:

$$g^{L-1} = 2\left(g^L + d^L\right) \quad \text{Eq. 7}$$

The similar optical flow estimation process (i.e. solution of Eq. 6) will be done for the level $L_m - 1$. The upsampling and optical flow will continue for the level $L = 0$. To this end, the final optical flow value is:

$$d = g^0 + d^0. \quad \text{Eq. 8}$$

**Motion trajectory analysis**

All key points in each frame are tracked and captured by the optical flow algorithm. According to the coordinates of the key points, the motion trajectory graph is generated by recording the time-resolved travel distance $D_n$ (n is the number of frames.) from the average coordinate to the image origin of each frame. As a result, the frame with the local maximum is ED and the one with the local minimum is ES, as illustrated in Figure 3.

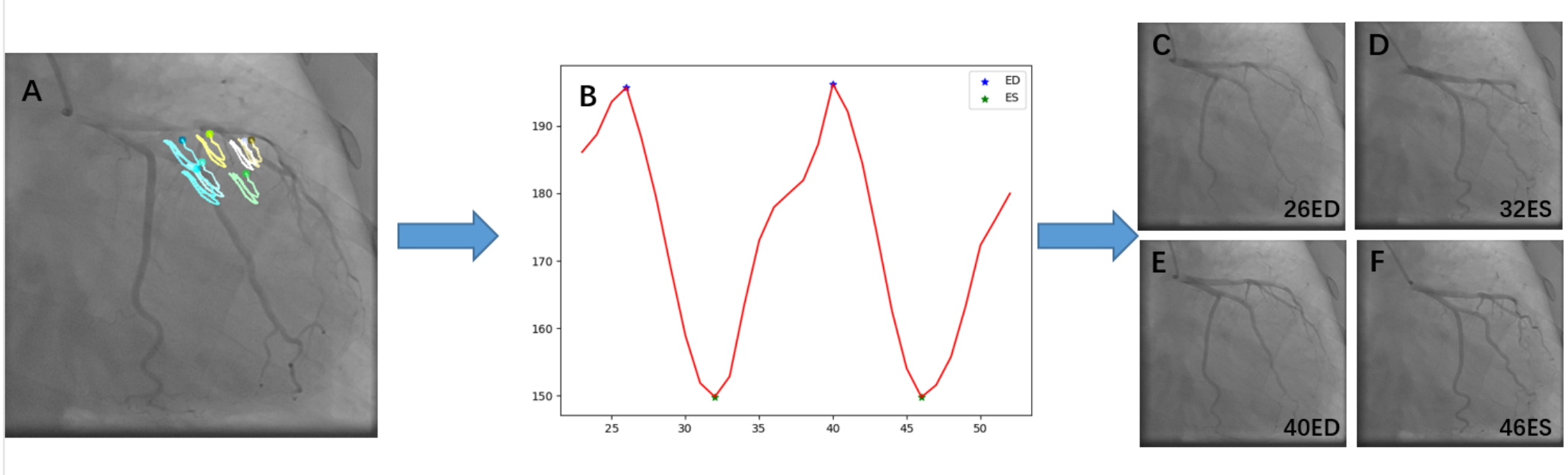

**Figure 3.** Automatic identification of the end-diastolic and end-systolic frames from the motion trajectories of key points. (A) The trajectory of the key points on the ICA frames. (B) Motion trajectory of $D_n$,where $n$ is the total number of frames. (C) and (E) are the resulting ED frames, (D) and (F) are the resulting ES frames.

## 3. Experiments and Results

### Experimental settings

A total of 224 images (ED: 114 frames, ES: 110 frames) were tested in 62 ICA videos. Two independent readers blinded to our algorithms and results visually identified ED and ES frames and then interpreted them together to generate a consensus identification. The consensus visual identification results were used as the ground truth. The results by our automated approach were compared with the ground truth for evaluation.

### Validation of our approach against the ground truth and individual readings

Table 2 shows the differences with 0-3 frames between our approach and visual identification results for data from two sites. For the combination of Site 1 and Site 2, when compared with the consensus readings, the percentages of frame error within a one-frame range were 92.86% for both ED and ES frames, 92.99% for ED frames alone, 92.73% for ES frames alone. For Site 1 only, the percentages of frame error within a one-frame range were 93.00% for both ED and ES frames, 94.59% for ED frames alone, 91.30% for ES frames alone; For Site 2 only, the percentages of differences with at maximum one frame were 92.59% for both ED and ES frames, 90.00% for ED frames alone, 95.12% for ES frames alone.

**Table 2**. The differences of end-diastolic (ED) and end-systolic (ES) frames between our approach and the visual identifications by Reader 1, Reader 2 and their consensus. DF0, DF1, DF2, and DF3 mean the frame differences of 0, 1, 2, and 3, respectively. Note that the maximum frame difference was 3.

| | Site 1 (Overall: 143, ED: 74, ES: 69.) | | | | Site 2 (Overall: 81, ED: 40, ES: 41.) | | | |
|---|---|---|---|---|---|---|---|---|
| | DF0 | DF1 | DF2 | DF3 | DF0 | DF1 | DF2 | DF3 |
| Reader 1 | | | | | | | | |
| ED | 51(68.92%) | 22(29.73%) | 1(1.35%) | 0(0.00%) | 19(47.50%) | 18(45.00%) | 2(5.00%) | 1(2.50%) |
| ES | 43(62.32%) | 21(30.43%) | 4(5.80%) | 1(1.45%) | 21(51.22%) | 17(41.46%) | 3(7.32%) | 0(0.00%) |
| Overall | 94(65.73%) | 43(30.07%) | 2(3.50%) | 1(0.70%) | 40(49.38%) | 35(43.21%) | 5(6.17%) | 1(1.23%) |
| Reader 2 | | | | | | | | |
| ED | 52(70.27%) | 17(22.97%) | 5(6.76%) | 0(0.00%) | 28(70.00%) | 8(20.00%) | 3(7.50%) | 1(2.50%) |
| ES | 38(55.07%) | 27(39.13%) | 4(5.80%) | 0(0.00%) | 29(70.73%) | 9(21.95%) | 3(7.32%) | 0(0.00%) |
| Overall | 90(62.43%) | 44(30.77%) | 9(6.29%) | 0(0.00%) | 57(70.37%) | 17(20.99%) | 6(7.41%) | 1(1.23%) |
| Consensus | | | | | | | | |
| ED | 54(72.97%) | 16(21.62%) | 4(5.14%) | 0(0.00%) | 22(55.00%) | 14(35.00%) | 2(5.00%) | 2(5.00%) |
| ES | 42(60.87%) | 21(30.43%) | 5(7.25%) | 1(1.45%) | 28(68.29%) | 11(26.83%) | 2(4.88%) | 0(0.00%) |
| Overall | 96(67.13%) | 37(25.87%) | 9(6.29%) | 1(0.70%) | 50(61.73%) | 25(30.86%) | 4(4.94%) | 2(2.47%) |

### Influence of different thresholds on the results of automatic identification

The number of key points varies when different thresholds are applied to the same image. The higher the threshold, the fewer the key points from the Shi-Tomasi detection algorithm produces, but the quality of the key points improves with the increase of the threshold. As illustrated in Figure 4A, when the threshold was 0.5, the number of key points was large, and the captured arterial territory was large correspondingly; when the threshold was 0.8 or 0.9, the number of key points was small, and only a small territory was covered.

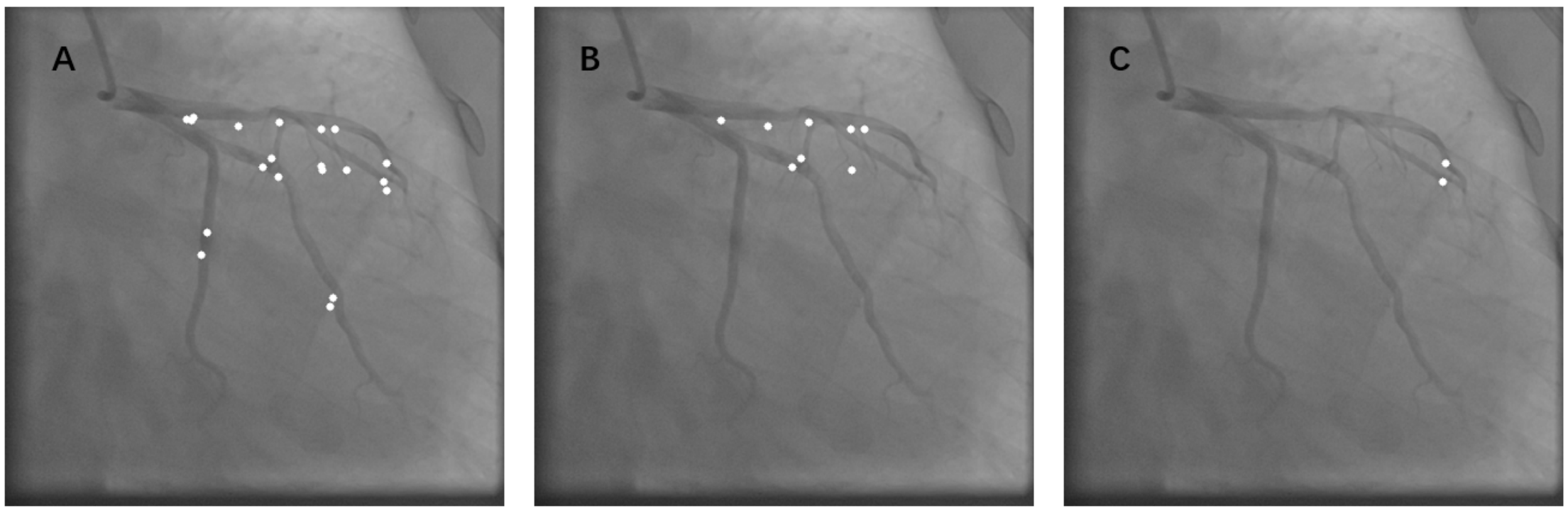

**Figure 4**. Key points captured by the Shi-Tomasi detection algorithm with different thresholds. (A) Threshold=0.5. (B) Threshold=0.8. (C) Threshold=0.9.

In order to find the optimal threshold, we performed three automatic identification experiments on the combination of Site 1 and Site 2 and then computed the percentages of differences based on consensus readings. Table 3 shows that when the threshold is 0.8, the percentage of differences with at maximum one is the largest. Thus, in our experiments, a threshold of 0.8 was used.

**Table 3.** The differences between automatic identification with three different thresholds for detecting key points and the consensus readings.

| Threshold \ Difference Frame | DF0 | DF1 | DF2 | DF3 |
|---|---|---|---|---|
| 0.5 | 138(61.61%) | 69(30.80%) | 15(6.70%) | 2(0.89%) |
| 0.8 | 146(65.18%) | 62(27.68%) | 13(5.80%) | 3(1.34%) |
| 0.9 | 126(56.25%) | 77(34.38%) | 20(8.93%) | 1(0.45%) |

**Inter-observer reproducibility during visual identifications**

Table 4 shows the difference between the results on 31 patients by two readers. For ED phase, the average difference between the two readers was 1.19 frames, and for the ES phase, the average difference was 1.71.

**Table 4**. The comparison of ED and ES visual reading between two readers.

| Difference Frame | Number of patients (ED) | Number of patients (ES) |
|---|---|---|
| 0 | 7 | 2 |
| 1 | 12 | 14 |
| 2 | 11 | 9 |
| 3 | 1 | 4 |
| 4 | 0 | 1 |
| 5 | 0 | 1 |
| Total difference | 37 | 53 |
| Average difference | $37/31 \approx 1.19$ | $53/31 \approx 1.71$ |

## 4. Discussion

A new method to automatically identify the time points within a cardiac cycle from ICA videos has been developed. Compared with the consensus results by two readers, the percentages of frame error within a one-frame range were 92.86% for both ED and ES frames, 92.99% for ED frame alone, and 92.73% for ES frame alone.

A number of researchers have developed techniques to detect the cardiac phase on cardiovascular images.

1) Echocardiographic images. Parisa et al. [21] applied a manifold learning method to two-dimensional echocardiographic images; a locally linear embedding (LLE) algorithm was used to represent the nonlinear embedded information in a sequence of images in a two-dimensional manifold, each image was represented by a point on the reconstructed manifold, and the distances between consecutive points in the manifold were used to identify the ED and ES frames. For the ED phase, the difference of 1.3 frames on average and a maximum of 3 frames, and for the ES phase, the average difference was 0.7 frames and a maximum of 2 frames. Tsui et al. [22] proposed an automatic systolic-diastolic classification method for real-time three-dimensional transesophageal echocardiography (RT-3D-TEE) data. This method performed threshold segmentation and median filtering to denoise the mitral valve and then classified the mitral valve based on the number of mitral valve regions. The classification accuracy of this method reached an accuracy of 91.04% for both ED and ES frames.

2) Cardiac magnetic resonance (CMR) images. Yang et al. [23] proposed a method to automatically identify ED and ES frames in free-breathing CMR imaging. The left ventricle (LV) was positioned using a convolutional neural network, the convex hull of the LV blood pool was processed, and then the least square method was used to perform ellipse fitting and center positioning of the LV. The signal curve generated by the center position of the LV was processed by a low-pass filter to obtain the respiratory motion signal. The minimum value of the normalized cross-correlation and the area fitted by the LV ellipse was used to identify ED and ES frames. The average accuracy of all slices is 76.5% for the ED and ES frames.

Nevertheless, the two-dimensional projection and low contrast of the myocardial wall make it difficult to detect the wall motion and report the ED/ ES frames in ICA videos. Ciusdel et al. [24] proposed a fully image-based method based on two deep neural networks to detect the cardiac phase and ED frames in ICA videos. The first deep neural network (DNN) was employed to detect coronary arteries and thus preselect a subset of frames in which coronary arteries are well visible. The second DNN used the ground truth provided by the ECG to predict the cardiac phrase of each frame. This method had a precision of 97.4% and a recall of 96.9% for the detection of ED frames. Although it has high accuracy, it required a large number of samples to train the model and the interpretability is limited by the DNN model. Our approach achieves excellent accuracy with a short time in two independent datasets. It tracks the motion trajectories of key points automatically selected in ICA videos and all the processing steps are clearly visualized and can be improved as a relatively independent module, thus representing a significant innovation to solve this unique problem.

**Factors that may influence the accuracy of our approach**

Two factors may affect the accuracy of automatic identification of ED and ES frames. Due to ventricular fibrillation in some patients, two consecutive maximum or minimum values may be present during the motion trajectory analysis of the key points of the blood vessels. In this situation, it is also difficult for clinicians to determine the ground truth. Moreover, our approach may fail due to extremely poor imaging quality or camera movement in the imaging process.

**Clinical applicability**

Automatic identification of ED and ES frames in ICA videos is important for the subsequent cardiovascular image analysis, particularly the 3D reconstruction of arterial anatomy [10], coronary flow quantification, and multi-modality fusion between ICA and myocardial images. Therefore, our approach has a great promise for clinical use.

The clinical verification with 31 patients confirmed the applicability of our automatic identification approach. There are high agreement rates between the frames by our approach and the doctor's consensus reading, ensuring the accuracy of the automatic identification. From clinical experts' perspectives, the identification error within a one-frame range is acceptable. Compared with the visual identification time of readers, the average processing time of 8 secs is relatively short, which further confirms the applicability of our approach.

The selection of key points plays an important role in the accuracy and reproducibility of the automatic identification of ED and ES frames. The only user involvement, drawing the arterial territory, may affect the reproducibility. The number and quality of key points from the Shi-Tomasi detection algorithm are influenced by the thresholds as illustrated in Figure 4. Our experiments in Table 3 show that a fixed threshold of 0.8 has both accuracy and robustness. Nevertheless, the key points may be selected with new algorithms. In [25],[26] we developed new methods to segment the entire arterial tree and extract individual arteries through semantic segmentation from fluoroscopy angiograms. The results from these segmentation methods will be used to automatically obtain the bifurcation points in the arterial tree. The user involvement in drawing the arterial territory will also be eliminated. Both the operating time and inter-variation would be further reduced if successful.

**Study Limitations**

Our approach was tested on a relatively small data set. In addition, the consensus readings interpreted by two readers were used as the ground truth. The electrocardiogram (ECG)-gated acquisition in ICA videos can show more accurate time points and identify ED and ES frames, and thus the ground truth can be more accurate.

## 5. Conclusions

A new approach has been developed and validated to identify ED and ES frames in ICA videos. Compared with the consensus results by two readers, the percentages of frame error within a one-frame range were 92.86% for both ED and ES frames, 92.99% for ED frame alone, and 92.73% for ES frame alone. It has great promise to assist in subsequent ICA image analysis.